\documentclass{rspublic}
\usepackage{graphicx}
\usepackage{natbib}
\bibpunct[, ]{(}{)}{;}{a}{}{,}
\begin{document}

\title[No SNe in two long GRBs]{No supernovae detected in two long-duration Gamma-Ray Bursts}

\author[D. Watson and others]{D.~Watson$^1$, J.~P.~U.~Fynbo$^1$,
C.~C.~Th\"one$^1$, J.~Sollerman$^{1,2}$}

\affiliation{$^1$Dark Cosmology Centre, Niels Bohr Institute, University of
Copenhagen, Juliane Maries Vej 30, DK-2100 Copenhagen \O, Denmark\\
$^2$Department of Astronomy, Stockholm University, Sweden}

\label{firstpage}

\maketitle

\begin{abstract}{gamma rays: bursts -- supernovae: general}

There is strong evidence that long duration gamma-ray bursts (GRBs) are 
produced during the collapse of a massive star. In the standard version of 
the Collapsar model, a broad-lined and luminous Type Ic 
core-collapse supernova (SN) accompanies the GRB. This association has 
been confirmed in
observations of several nearby GRBs. Recent observations show that
some long duration GRBs are different. No SN emission
accompanied the long duration GRBs 060505 and 060614 down to limits
fainter than any known Type Ic SN and hundreds of times fainter than the archetypal SN\,1998bw that 
accompanied GRB\,980425.
Multi-band observations of the early afterglows, as well as spectroscopy of
the host galaxies, exclude the possibility of significant dust obscuration.
Furthermore, the bursts originated in star-forming galaxies, and in
the case of GRBs\,060505 the burst was localised to a compact 
star-forming knot in a spiral arm of its host galaxy. We find that the
properties of the host galaxies, the long duration of the bursts and, in the
case of GRB\,060505 the location of the burst within its host, all imply
a massive stellar origin. The absence of a SN to such deep limits 
therefore suggests a new phenomenological type of massive stellar death.  

\end{abstract}

\section{Introduction}
The collapse of massive stars on some occasions produce long duration
$\gamma$-ray bursts (GRBs). It was a common expectation that broad-lined and
luminous Type Ic core-collapse supernovae (SN) accompany every long-duration
GRB \citep{2004ApJ...609..952Z,1998Natur.395..670G,2003Natur.423..847H,2003ApJ...591L..17S,2004ApJ...609L...5M,2006Natur.442.1011P},
though it was not clear that such an expectation was justified on theoretical grounds.

The GRBs 060505 and 060614 were detected by (the dedicated GRB satellite)
\emph{Swift}'s $\gamma$-ray Burst Alert
Telescope (BAT) on 2006 May 5.275 and 2006 June 14.530 respectively
\citep{2006GCN..5142....1H,2006GCN..5252....1P}. GRB\,060505 was a faint
burst with a duration of 4\3s. GRB\,060614 had a duration of 102\3s and a
pronounced hard to soft evolution. Both were rapidly localised by
\emph{Swift}'s X-ray telescope (XRT). Subsequent follow-up of these bursts led to the
discovery of their optical afterglows, locating them in galaxies at low
redshift: GRB\,060505 at $z=0.08913$ \citep{fynbo2006} and GRB\,060614 at
$z=0.12514$ \citep{2006astro.ph..8322D}. The relative proximity of these bursts engendered an
expectation that a bright SN would be discovered a few days after the
burst, as had been found before in other low-redshift long duration GRBs.

\section{Observations and results}

\begin{figure}
 \includegraphics[width=0.5\columnwidth]{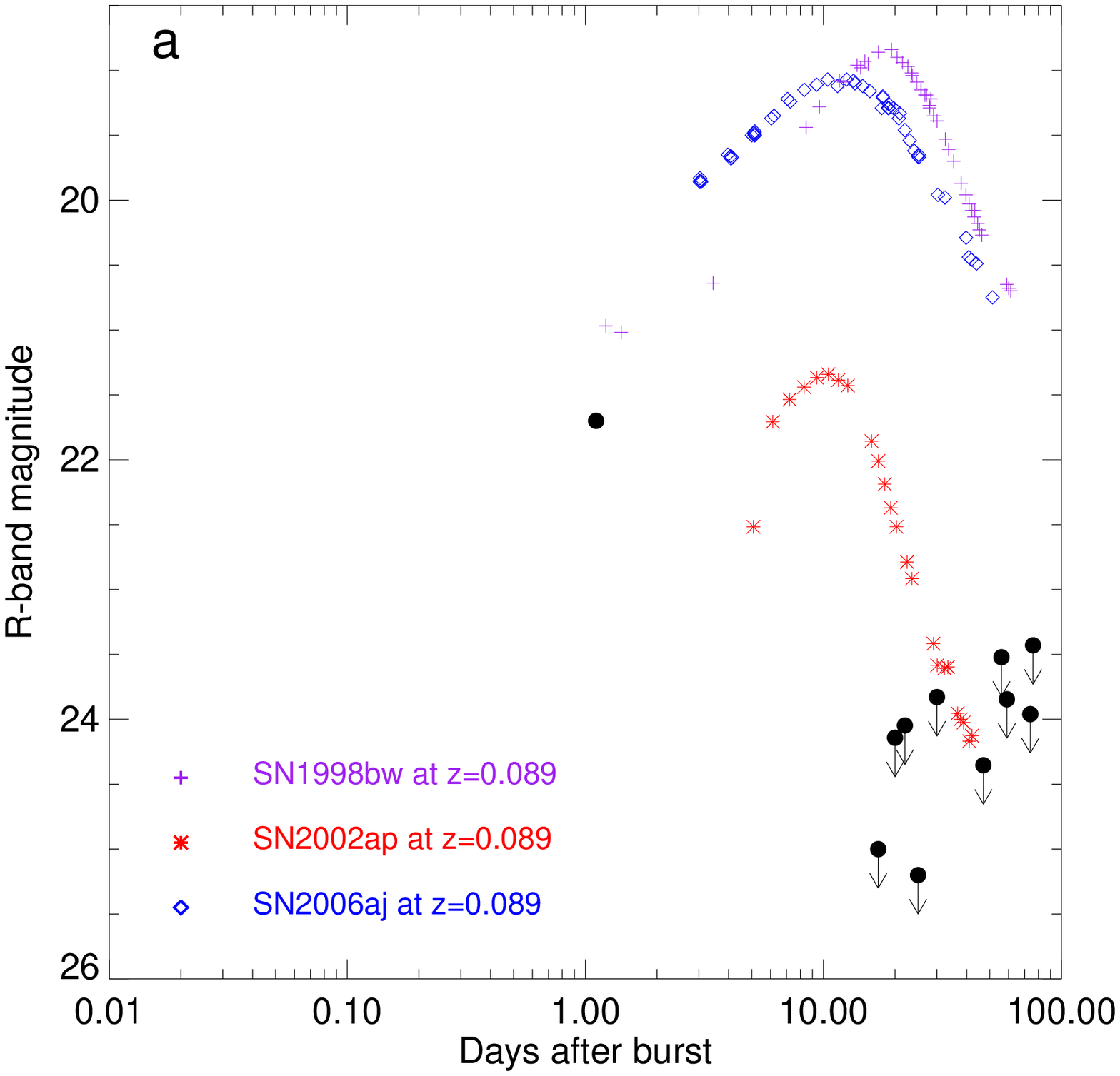}
 \includegraphics[width=0.5\columnwidth]{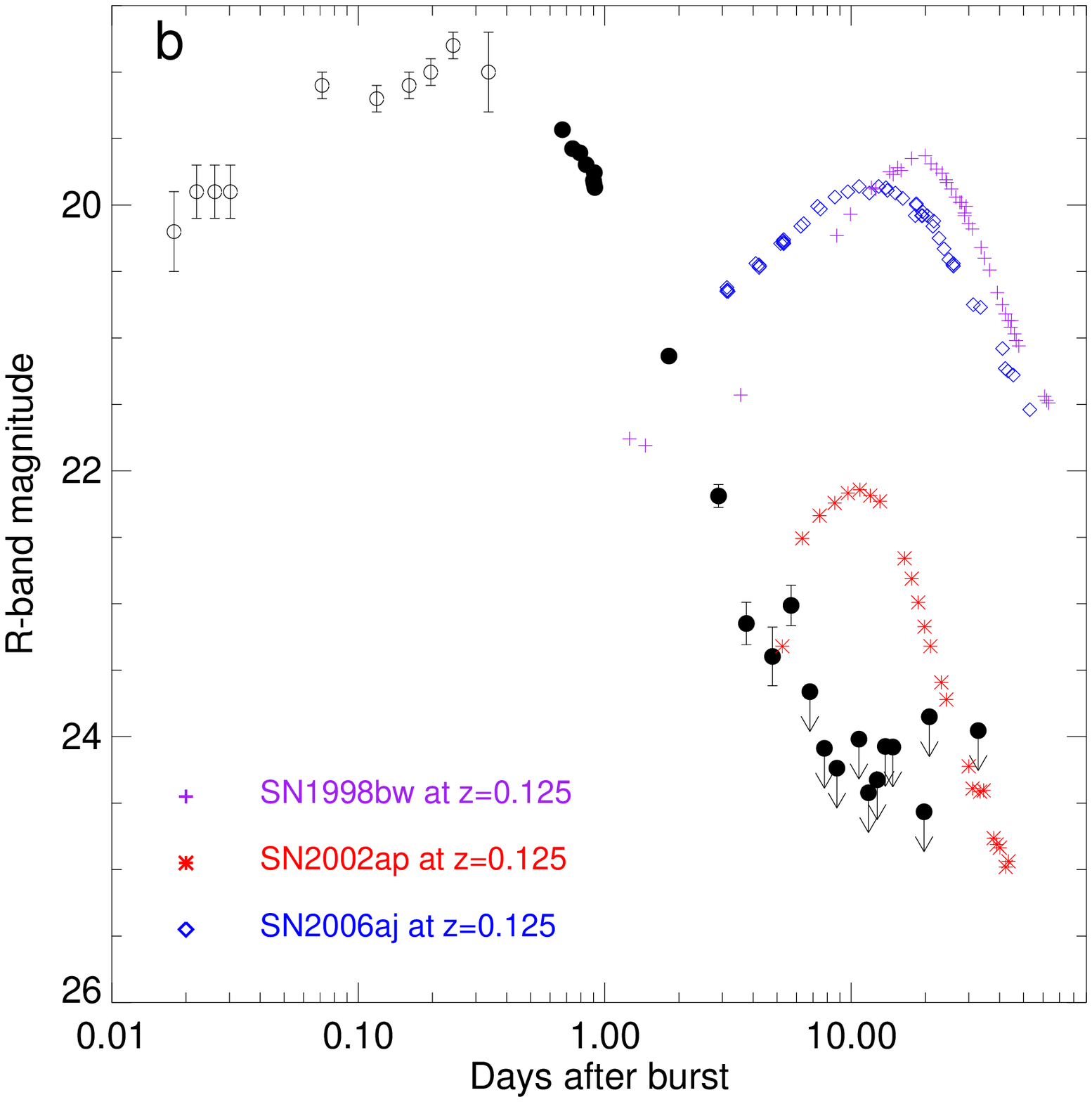}
 \caption{Optical afterglow light-curves of two nearby $\gamma$-ray bursts:
          a) GRB\, 060505 and b) GRB\,060614. The light-curves
          of the Ic SNe 1998bw, 2002ap and 2006aj, are plotted as they would have appeared at
          the redshift of GRB\,060505 (a, left) and at the redshift of
          GRB\,060614 (b, right). Afterglow detections (black, filled circles) and
          subsequent $3\sigma$ upper limits are plotted. We conclude
          that neither GRB\,060505 nor GRB\,060614 were associated with
          significant SN emission down to very faint limits hundreds of
          times less luminous than the archetypal SN\,1998bw.
          For GRB\,060614 the early light-curve data
          is taken from \citet{2006GCN..5255....1H} (open
          circles).}
 \label{fig:lightcurve}
\end{figure}

We monitored the afterglows of GRB\,060505 and 060614 at
optical/near-infrared wavelengths. This led to early
detections of the afterglows. We continued the monitoring campaign and
obtained stringent upper limits on any re-brightening at the position of the
optical afterglows up to 12 and 5 weeks after the bursts, respectively. The
light-curves obtained based on this monitoring are shown in
figure~\ref{fig:lightcurve}. For GRB\,060505 we detected the optical
afterglow
at a single epoch in multiple bands ($B$, $V$, $R$ and $I$, Xu et~al.\ in
preparation). All subsequent
observations resulted in deep upper limits. For GRB\,060614 we followed the
decay of the optical afterglow in the R-band up to four nights after the
burst. No source was detected to deep limits at later times. As seen in
figure~\ref{fig:lightcurve}, the upper limits are far below the fluxes of
previous Ic SNe, in particular SNe associated with long GRBs
\citep{1998Natur.395..670G,2003Natur.423..847H,2003ApJ...591L..17S,2004ApJ...609L...5M,2006Natur.442.1011P}.
For both GRBs our $3\sigma$ limit around the time of expected maximum of a
SN component is 80--100 times fainter
than SN\,1998bw. The very deep observations of GRB\,060505 on May
23 and May 30 give a $3\sigma$ upper limit more than 250 times fainter
than SN\,1998bw at a similar time. Hence, any SN associated with GRB\,060505
must have had a peak magnitude in the $R$-band fainter than about $-13.5$.  

These two SN-less GRBs share no obvious characteristics in
their prompt emission. GRB\,060505 was one of the least luminous
bursts discovered with Swift with an isotropic-equivalent energy
release of $1.2\times10^{49}$\3erg. It had a relatively short duration, and
was single-peaked with a very faint afterglow. GRB\,060614 was
about a hundred times more luminous with an isotropic-equivalent
energy release of $8.9\times10^{50}$\3erg, and it showed strong spectral
evolution. Its optical afterglow brightened for the first
day \citep{2006GCN..5258....1S}, reminiscent of GRB\,970508
\citep{1998ApJ...496..311P}.

\subsection{Low extinction}
An obvious explanation to explore for the absence of a SN is the 
presence of dust along the line of sight.
In these cases however, we are fortunate that the levels of Galactic extinction
in both directions are very low, E($B-V$)$ = 0.0216$\,mag. In the case of GRB\,060505,
our spatially resolved
spectroscopy of the host galaxy allows us to use the Balmer emission line
ratios to limit the dust obscuration at the location of the burst. The
Balmer line ratio is consistent with no internal reddening. In the case of
GRB\,060614, the detection of the early afterglow in many bands, including
the \emph{Swift} UV bands UVW1 and UVW2 \citep{2006GCN..5255....1H}, rules out
significant obscuration of the
source in the host galaxy and we conclude that there is no significant dust
obscuration in either case \citep[see also][]{2006astro.ph..8322D}.

\subsection{Host galaxies}

\begin{figure}
 \includegraphics[width=\columnwidth,clip=]{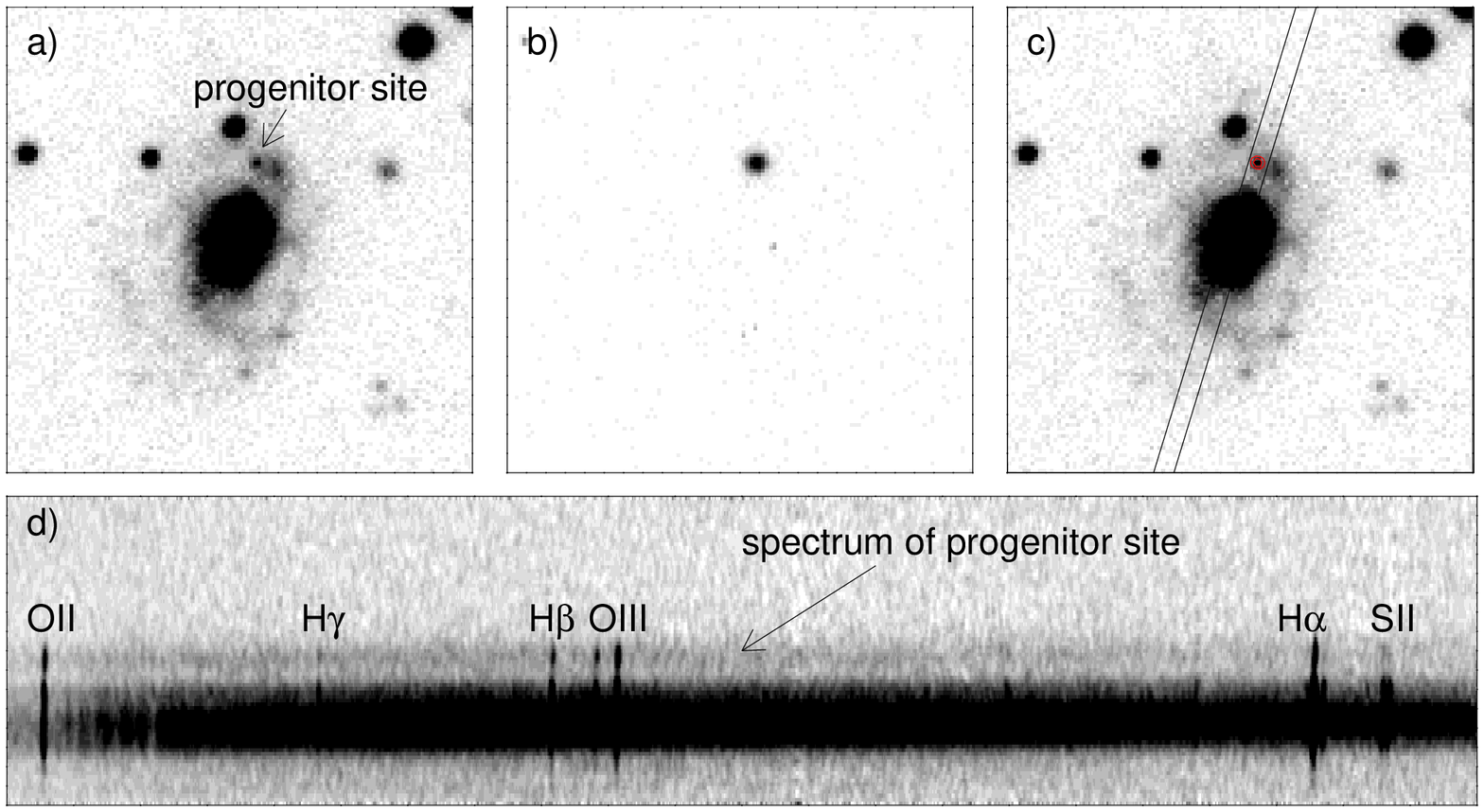}
 \includegraphics[width=\columnwidth]{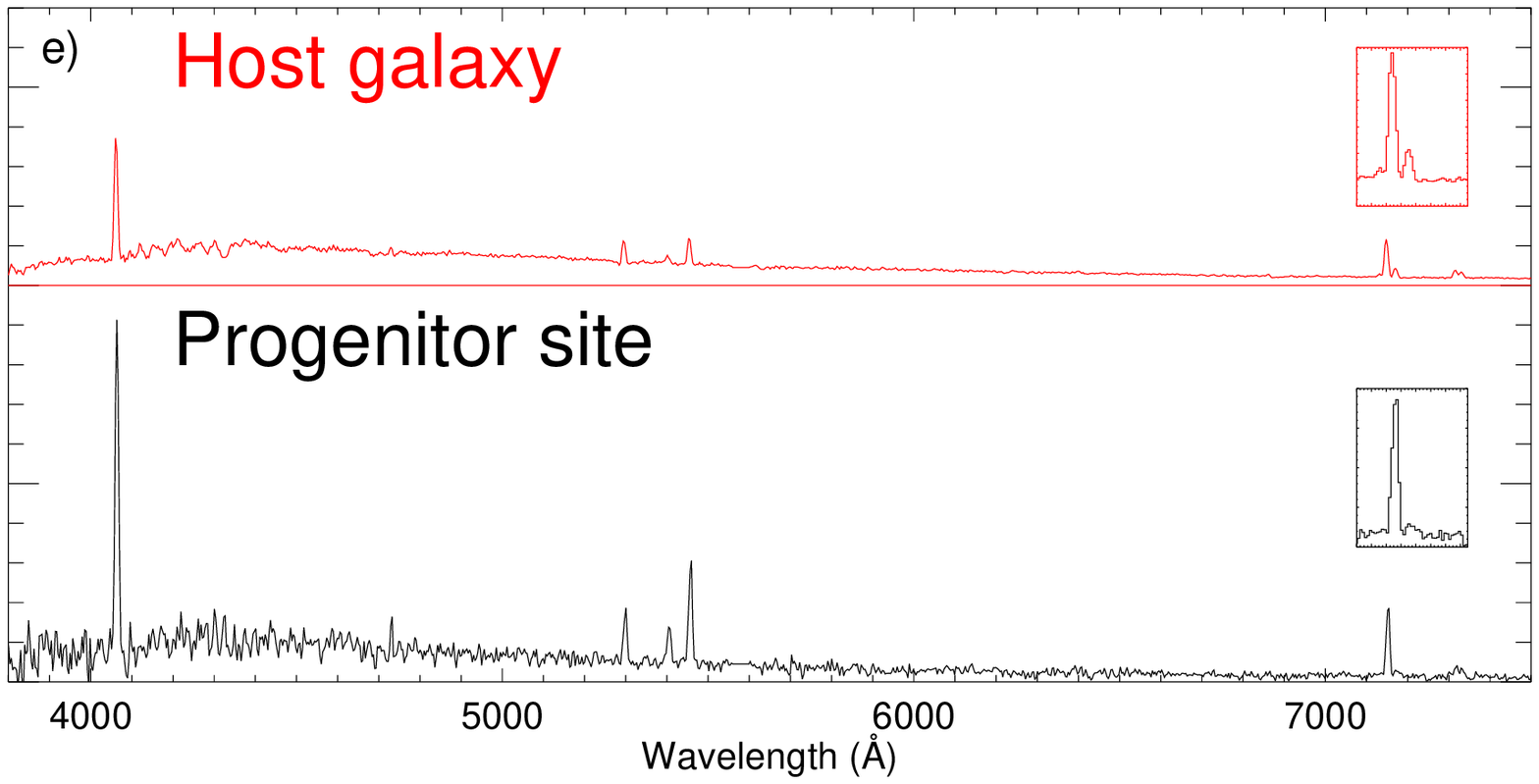}
 \caption{VLT images and spectra of the GRB\,060505 host galaxy. a) The
          field ($24^{\prime\prime}\times 24^{\prime\prime}$) of GRB 060505 as observed from the VLT
          in the R- band on Sep 14. The arrow marks the position where the
          optical afterglow was detected. The source seen at this position
          in the image is a compact star-forming region in which the
          progenitor of GRB\,060505 was located. b) The result of subtracting
          the Sep 14 image seen in a from the May 6 image. Seen is the
          optical afterglow component alone. c) As (a), but with the position
          of the optical afterglow marked with red contours and with the
          orientation of the slit for the May 23 spectrum indicated. The
          position of the afterglow is within the astrometric uncertainty of
          less than $0.05^{\prime\prime}$ coincident with the position of the compact
          star-forming region. d) The 2-dimensional optical spectrum
          obtained with VLT/FORS2 on May 23. As seen in (c) the slit covered
          the centre of the host galaxy and the location of GRB 060505. As
          seen in the spectrum, this site is indeed a bright star-forming
          region in the host galaxy and we hence have very strong evidence
          that the 060505 progenitor was a massive star. From the ratio of
          the Balmer line strengths we exclude dust extinction of more than
          a few tenths of a magnitude in the R-band.
          e) Spectra of the host galaxy (upper pane) and star-formation knot
          (lower pane) associated with GRB\,060505. Hydrogen and oxygen
          lines associated with star-formation are clearly detected.
          \emph{Insets}: H$\alpha$ and [N\,\textsc{ii}]. The ratio of
          [N\,\textsc{ii}]\,$\lambda$/H$\alpha$ is smaller in the
          spectrum of the progenitor site, implying a lower metallicity
          and/or a more intense ionising flux here than in the host as a whole.}
 \label{fig:2dspec}\label{fig:hostspec}
\end{figure}

Both GRBs were located in star-forming galaxies.
The host galaxy of GRB\,060505 has an absolute
magnitude of about M$_B = -19.6$ and the spectrum displays the prominent
emission lines typically seen in star-forming galaxies. The 2-dimensional
spectrum shows that the host galaxy emission seen at the position of the
afterglow is due to a compact H\,\textsc{ii} region in a spiral arm of the host (see
figure.~\ref{fig:hostspec} and Th\"one et~al.\ in prep.). We estimate a star-formation rate (SFR)
of 1\3M$_\odot$\3yr$^{-1}$ and a specific SFR of about 4\3M$_\odot$\3yr$^{-1}$\3(L/L*)$^{-1}$ (assuming
M*$_B = -21$).

The host galaxy of GRB\,060614 is significantly fainter than the host of
GRB\,060505, with an absolute
magnitude of about M$_B = -15.3$. This is one of the least luminous GRB host
galaxies ever detected. GRB\,060614 was observed with the GMOS imaging 
spectrometer on the Gemini 8-m on the night of 29 July 2006. We detected the H and
[O\,\textsc{iii}] emission lines and infer an SFR of
0.014\3M$_\odot$\3yr$^{-1}$. The specific SFR is 3\3M$_\odot$\3yr$^{-1}$\3(L/L*)$^{-1}$.

Sub-L*, star-forming host galaxies, like these two, are fairly common among long GRB
host galaxies \citep{2006Natur.441..463F}. For comparison, the specific SFRs
of the 4 previously studied nearby ($z<0.2$) long GRB host galaxies are 6, 7, 25 and
39\3M$_\odot$\3yr$^{-1}$\3(L/L*)$^{-1}$ \citep{2005NewA...11..103S,2006A&A...454..503S}.
Hence, the specific SFRs of the two hosts studied here are
slightly lower than the four previously studied nearby hosts, but the scatter
in the rates is large and it is not possible to conclude 
that the host galaxies of GRB\,060505 and GRB\,060614 are qualitatively different
from other $z<0.2$ long-GRB host galaxies.

\section{Could GRB\,060505 and GRB\,060614 be short GRBs or at higher redshifts?}

The duration of the prompt emission GRB\,060505 and GRB\,060614 is 4\3s and
102\3s, respectively. Hence, they are both well outside the classification of
short bursts, i.e.\ duration less than 2\3s \citep{1993ApJ...413L.101K}. It
is worth noting in this context that the typical long
GRBs 000301C at $z=2.04$ and 020602 at $z=4.05$ had durations of only 2\3s and
7\3s, respectively, so their rest-frame durations are significantly shorter
than the 4\3s of even GRB\,060505. Short GRBs have previously had
accompanying SNe excluded \citep{2005ApJ...630L.117H,2006ApJ...638..354B}.
Therefore, it might be speculated that both
SN-less GRBs studied here were extreme members of the class of
progenitors responsible for short GRBs. Short GRBs have in some cases been
found to be associated with older stellar populations than long GRBs, and it
is widely expected that they are predominantly caused by merging compact
objects \citep{2005Natur.437..851G}.

In addition to their long duration, the facts that 1)
GRB\,060505 and GRB\,060614 can be localized to star-forming galaxies, 2) in the case of
GRB\,060505 even a star-forming region in a spiral arm of its host galaxy (see
figure~\ref{fig:hostspec}), and 3) in the case of GRB\,060614 the afterglow is located around
the half-light radius of its star-forming host \citep{2006astro.ph..8257G}
all strongly suggest that the
progenitors were massive stars. The evidence in this paper
should cause us to be more open-minded about the origins of short GRBs: absence
of SNe has been an argument used to suport the idea that short GRBs do not
have a massive stellar progenitor. Now we have observed long-duration SN-less GRBs in star-forming regions,
suggesting that a non-detection of a SN does not preclude a massive
progenitor. In the near future, the location of the GRB, i.e.\ in a star-forming region or in an
older component, may be the only way to discriminate between merging compact
objects and massive stars as progenitors. In fact, several host galaxies for
short GRBs have been found to be as actively star-forming as some host
galaxies of long GRBs \citep{2005Natur.437..859H,2006astro.ph..1455S}. The GRB labels `long' and `short' have become
synonymous with `massive stars' and `other progenitors'.  These distinctions
may need to be relaxed.

It has also been suggested that GRB\,060505 and
GRB\,060614 could in fact be more distant bursts and that the proposed hosts are
only foreground galaxies. In the case of GRB\,060505, it is very unlikely that
the afterglow is superposed precisely on a star-forming knot in the spiral arm of a
foreground galaxy (figure~\ref{fig:hostspec}c) if it was in fact more distant. In the
case of GRB\,060614 the impact parameter is about $0.5^{\prime\prime}$, and the afterglow is located
around the half-light radius of the proposed host galaxy \citep{2006astro.ph..8257G}. The
\emph{Swift}-UVOT detection of the afterglow \citep{2006GCN..5255....1H} places an upper
limit on the redshift of about $z\sim1$. At such relatively low redshift the host galaxy
should be detectable in the very deep HST images of the field, but no other galaxy than the proposed
host galaxy is seen \citep{2006astro.ph..8257G}. Furthermore, even at
$z\sim1.1$, any SN would still have to be $\gtrsim2$ magnitudes fainter than
SN\,1998bw to remain undetected. Therefore, it is very unlikely that either
GRB\,060505 or GRB\,060614 are at higher redshifts than $z=0.089$ and $z=0.125$, respectively.

\section{No detected SNe}
All spectroscopically confirmed SN-GRBs have peak
magnitudes within about half a magnitude of SN\,1998bw
\citep{2004ApJ...609..952Z,2006ARA&A..44..507W}. For X-ray Flashes
(XRFs) there has been some evidence
that associated SNe may span a somewhat wider range of
luminosities \citep{2006Natur.442.1011P,2005ApJ...622..977L,2005ApJ...627..877S,2004ApJ...609..962F}, but still
well within the range of non--GRB-selected Type Ic SNe.
Any SN associated with these two long GRBs must therefore have been
substantially fainter than any SN Ic seen to date. Scaling the peak magnitude
of a type Ic SN with the ejected $^{56}$Ni mass suggests that the mass of
$^{56}$Ni ejected in these cases is less than one percent of a solar mass.

The non-appearance of a SN in these cases was a surprise and indicates that we have
uncovered GRBs with quite different properties from those studied previously.
It is plausible that the origin of these bursts lies in one of the many SN-less
GRB progenitors suggested prior to the definitive association between GRBs and
SNe. We note, however, that our results confirm a prediction within the
Collapsar model: that while direct black hole (BH) formation is likely to
result in SNe associated with GRBs, `fallback'-formed BHs could produce SN-less GRBs
\citep{2006ApJ...650.1028F,2003ApJ...591..288H,2005NuPhA.758..263N}.
These authors cited immediately above argue that the nucleosynthetic yields
from the explosion related to
the fallback-formed BH are very different from those resulting from a
direct-formed BH. In the fallback case, little $^{56}$Ni is synthesized and
expelled, resulting in the prediction that only very faint SNe, perhaps
hundreds of times fainter than SN\,1998bw, should be observed in some GRBs.
Among core collapse SNe, a few very nickel deficient \citep{1998ApJ...493..933S}
and low velocity \citep{2003MNRAS.338..711Z} type
II SNe have been detected that were explained using this fallback mechanism 
\citep[see also]{2004MNRAS.347...74P}.
It has been pointed out that given the relatively less massive
progenitors, fallback BH formation should also be relatively common among
(Collapsar) GRBs. In another variant of the Collapsar model, progenitor stars
with relatively low angular momentum could produce SN-less GRBs
\citep{2003AIPC..662..202M}.

That some massive stars explode with a SN expelling large amounts of nucleosynthesised
gas at high velocities, while other massive stars die with a whisper, clearly
has consequences for our understanding of the energy input and the metal
enrichment to the interstellar medium. Until the era of gravitational wave
detectors \citep{2004PhRvD..69d4007V} or more sensitive neutrino astronomy
\citep{2005astro.ph..9570I}, it seems that GRBs will be the only 
signals we get from the deaths of some massive stars.
Of the six long GRBs or XRFs known to be at low redshift ($z < 0.2$), two now have no
associated SN, so the fraction of SN-less GRBs could be substantial.

\begin{acknowledgements}
We acknowledge benefits from collaboration within the EU
FP5 Research Training Network Gamma-Ray Bursts: An Enigma and a Tool. The
Dark Cosmology Centre is funded by the DNRF. The observations presented in
this article has been obtained from the ESO La Silla-Paranal observatory and
from the Gemini Observatory.
\end{acknowledgements}

\end{document}